\documentclass[journal]{IEEEtran}
\usepackage{amsmath}
\usepackage{graphicx}
\usepackage{caption2}
\usepackage{amsthm}
\usepackage{float}
\usepackage{mathrsfs}
\usepackage{verbatim}
\usepackage{epstopdf}
\usepackage{amssymb}
\usepackage{amsfonts}
\usepackage{subfigure}
\usepackage{color}
\usepackage{cite}
\usepackage{cancel}
\usepackage{amsmath,amssymb,amsthm,mathrsfs,amsfonts,dsfont}
\usepackage[shortlabels]{enumitem}

\DeclareMathOperator*{\argmax}{arg\,max}

\begin{document}
\title{Optimal Utility-Privacy Trade-off with \\Total Variation Distance as a Privacy Measure}
\author{\IEEEauthorblockN{
Borzoo Rassouli\IEEEauthorrefmark{2} and Deniz G\"{u}nd\"{u}z \IEEEauthorrefmark{3}\\}
\IEEEauthorblockA{ \IEEEauthorrefmark{2} University of Essex, b.rassouli@essex.ac.uk\\}
\IEEEauthorblockA{ \IEEEauthorrefmark{3} Imperial College London, d.gunduz@imperial.ac.uk}
\thanks{Most of this work was carried out when the first author was with the \textit{Information Processing and Communications Lab} at Imperial College London. This research was supported by the European Research Council (ERC) through the Starting Grant BEACON (agreement 677854), and by the UK Engineering and Physical Sciences Research Council (EPSRC) through the project COPES (EP/N021738/1).}}

\maketitle
\begin{abstract}
The total variation distance is proposed as a privacy measure in an information disclosure scenario when the goal is to reveal some information about available data in return of utility, while retaining the privacy of certain sensitive latent variables from the legitimate receiver. The total variation distance is introduced as a measure of privacy-leakage by showing that: i) it satisfies the post-processing and linkage inequalities, which makes it consistent with an intuitive notion of a privacy measure; ii) the optimal utility-privacy trade-off can be solved through a standard linear program when total variation distance is employed as the privacy measure; iii) it provides a bound on the privacy-leakage measured by mutual information, maximal leakage, or the improvement in an inference attack with a bounded cost function. 
\end{abstract}
\vspace*{-0.5cm}
\section{Introduction}

We measure, store, and share an immense amount of data about ourselves, from our vital signals to our energy consumption profile. We often disclose these data in return of various services, e.g., better health monitoring, a more reliable energy grid, etc. However, with the advances in machine learning techniques, the data we share can be used to infer more accurate and detailed personal information, beyond what we are willing to share. One solution to this problem is to develop privacy-preserving data release mechanisms that can provide a trade-off between the utility we receive and the information we leak. Denoting the data to be released by random variable $Y$, and the latent private variable as $X$, we apply a \textit{privacy-preserving mapping} on $Y$, whereby a distorted version of $Y$, denoted by $U$, is shared instead of $Y$. Typically, privacy and utility are competing goals: The more distorted version of $Y$ is revealed, the less information can be inferred about $X$, while the less utility can be obtained. As a result, there is a trade-off between obtaining utility and leaking privacy.

{\color{black}Since privacy can be a concern in legal transactions of data, it appears in different areas, where information is transferred from a user to a legitimate receiver of information. For instance, in database privacy \cite{Rebollo,Rebollo2, Sankar3}, data is published publicly, while preserving the privacy of individuals (identity, attributes, etc.). Another example is privacy in smart grids \cite{Sankar5, Khisti2, Tan, Giulio}, where a smart meter measures and reports the power consumption of a user to the electricity provider to improve the reliability and energy efficiency, and from this information, several private features of the user, such as their usage patterns or daily life habits, can be leaked.} 

The statistical view of privacy (information-theoretic, estimation-theoretic, and so on) has gained increasing attention recently \cite{Calmon1,Makhdoumi, Shahab5, As1,Ishwar, Shahab6, Huang1, SA2, Ish22}. For example, in \cite{Calmon1}, a general statistical inference framework is proposed to capture the loss of privacy in legitimate transactions of data. In \cite{Makhdoumi}, the privacy-utility trade-off under the log-loss cost function is considered, called the \textit{privacy funnel}, which is closely related to the \textit{information bottleneck} introduced in \cite{Tishby}. In \cite{Shahab5, As1}, the privacy and utility are expressed in terms of correctly guessing probabilities. In \cite{Ishwar}, a generic privacy model is considered, where the privacy mapping has access to a noisy observation $W$ of the pair $(X,Y)$. Different well-known privacy measures and their characteristics are also investigated in \cite{Ishwar}. 

We study the information-theoretic privacy in this paper. {\color{black}For two probability mass function $p,q$ on random variable $\mathcal{X}$, the total variation distance is defined as
\begin{equation}
\delta\bigg(p_X(\cdot),q_X(\cdot)\bigg)\triangleq \frac{1}{2}\|\mathbf{p}-\mathbf{q}\|_{1},
\end{equation}
where $\mathbf{p}$ and $\mathbf{q}$ are the probability vectors corresponding to probability mass functions (pmf) $p_{X}(\cdot)$ and $q_X(\cdot)$, respectively.
We measure the privacy-leakage (about the private variable $X$ by revealing $U$) by the following {\color{black}average total variation distance }
\begin{align}\label{pr}
T(X;U)&\triangleq\mathds{E}_U\bigg[\delta\bigg(p_{X|U}(\cdot|U),p_X(\cdot)\bigg)\bigg]\nonumber\\&=\frac{1}{2}\sum_{u}p_U(u)\|\mathbf{p}_{X|u}-\mathbf{p}_X\|_{1}.
\end{align}}
Note that $T$ is not symmetric, and we have $T(X;U)=0$ iff $X$ and $U$ are independent.

First, we characterize the optimal utility-privacy trade-off under this privacy measure for three different utility measures, namely mutual information, minimum mean-square error (MMSE), and probability of error. Then, we motivate the proposed privacy measure by showing that it satisfies both the \textit{post-processing} and \textit{linkage} inequalities \cite{Ishwar}, and it provides a bound on the leakage measured by mutual information, maximal leakage, or the improvement in an inference attack with an arbitrary bounded cost function as considered in \cite{Calmon1}. 

\textbf{Notations.} Random variables are denoted by capital letters, their realizations by lower case letters. Matrices and vectors are denoted by bold capital and bold lower case letters, respectively. 
For integers $m\leq n$, we have the discrete interval $[m:n]\triangleq\{m, m+1,\ldots,n\}$. 
For an integer $n\geq 1$, $\mathbf{1}_n$ denotes an $n$-dimensional all-one column vector. For a random variable $X\in\mathcal{X}$, with finite $|\mathcal{X}|$, the probability simplex $\mathcal{P}(\mathcal{X})$ is the standard $(|\mathcal{X}|-1)$-simplex given by
\begin{equation*}
\mathcal{P}(\mathcal{X})=\bigg\{\mathbf{v}\in\mathbb{R}^{|\mathcal{X}|}\bigg|\mathbf{1}_{|\mathcal{X}|}^T\cdot\mathbf{v}=1,\ v_i\geq 0,\ \forall i\in [1:|\mathcal{X}|]\bigg\}.
\end{equation*}
Furthermore, to each pmf on $X$, denoted by $p_{X}(\cdot)$, corresponds a probability vector $\mathbf{p}_X\in \mathcal{P}(\mathcal{X})$, whose  $i$-th element is $p_X(x_i)$ ($i\in[1:|\mathcal{X}|]$). Likewise, for a pair of random variables $(X,Y)$ with joint pmf $p_{X,Y}$, the probability vector $\mathbf{p}_{X|y}$ corresponds to the conditional pmf $p_{X|Y}(\cdot|y),\forall y\in\mathcal{Y}$, and $\mathbf{P}_{X|Y}$ is an $|\mathcal{X}|\times|\mathcal{Y}|$ matrix with columns $\mathbf{p}_{X|y},\forall y\in\mathcal{Y}$.
$F_{Y}(\cdot)$ denotes the cumulative distribution function (CDF) of random variable $Y$. For $0\leq t\leq 1$, $H_b(t)\triangleq-t\log_2t-(1-t)\log_2(1-t)$ denotes the binary entropy function with the convention $0\log0=0$. Throughout the paper, for a random variable $Y$ with the corresponding probability vector $\mathbf{p}_Y$, the entropies $H(Y)$ and $H(\mathbf{p}_Y)$ are written interchangeably. For $\mathbf{x}\in\mathbb{R}^n$ and $p\in[1,\infty]$, the $L^p$-norm is defined as $\|\mathbf{x}\|_p\triangleq(\sum_{i=1}^n|x_i|^p)^{\frac{1}{p}},p\in[1,\infty)$, and $\|\mathbf{x}\|_\infty\triangleq\max_{i\in[1:n]}|x_i|$. {\color{black} Let $p,q$ be two arbitrary pmfs on $X$. The Kullback–Leibler divergence from $q$ to $p$ is defined as $D(p||q)\triangleq\sum_xp(x)\log_2(\frac{p(x)}{q(x)})$.}
 
\section{System model and preliminaries}
Consider a pair of random variables $(X,Y)\in\mathcal{X}\times\mathcal{Y}$ ($|\mathcal{X}|,|\mathcal{Y}|<\infty$) distributed according to the joint distribution $p_{X,Y}$. We assume that $p_Y(y)>0,\forall y\in\mathcal{Y}$, and $p_X(x)>0,\forall x\in\mathcal{X}$, since otherwise the supports $\mathcal{Y}$ or/and $\mathcal{X}$ could have been modified accordingly. Let $Y$ denote the available data to be released, while $X$ denote the latent private data.
Assume that the \textit{privacy mapping}/\textit{data release mechanism} takes $Y$ as input and maps it to the \textit{released data} denoted by $U$. In this scenario, $X-Y-U$ form a Markov chain, and the privacy mapping is denoted by the conditional distribution $p_{U|Y}$. Let $J(X;U)\in[0,+\infty)$ be a generic privacy measure as a functional of the joint distribution $p_{X,U}$ that captures the amount of (information) leakage from $X$ to $U$. Hence, the smaller $J(X;U)$ is, the higher privacy is achieved by the mapping $p_{U|Y}$. Also, let $R(Y;U)\in[0,+\infty)$, a functional of the joint distribution $p_{Y,U}$, denote an application-specific quantity that measures the amount of utility/reward obtained by disclosing $U$. Therefore, the utility-privacy trade-off can be written as
\begin{equation}\label{toff}
\sup_{\substack{p_{U|Y}:\\X-Y-U\\J(X;U)\leq\epsilon}}R(Y;U),
\end{equation}
where the task is to find a privacy mapping that maximizes the utility, while guaranteeing a privacy-leakage up to the level $\epsilon$.

{\color{black}Having $p_{X|U}(\cdot|u)\neq p_{X}(\cdot)$ for some $u\in\mathcal{U}$, makes the private data potentially at risk. In other words, the adversary may gain some information about the private data due to this statistical dependence. Therefore, a measure of the distance between the posterior and the prior distributions of the private data can be adopted as a privacy measure. For example, the mutual information, i.e., $I(X;U)$, is the average Kullback-Leibler distance from $p_X$ to $p_{X|U}$, where the averaging is over the realizations of $U$.
In this paper, we use {\color{black}the average total variation distance between $p_{X|U}(\cdot|u)$ and $p_X(\cdot)$} to measure the privacy-leakage\footnote{In \cite{Bori}, the maximum total variation distance, where the maximum is over the realizations of $U$, is employed as the privacy-leakage measure.} as in (\ref{pr}), i.e., $J(X;U)=T(X;U)$. The adoption of this privacy measure is justified in the subsequent sections.}

{\color{black}Throughout the paper, we will refer to three other privacy measures, which are introduced next. 
The \textit{maximal leakage} \cite{Issa} from $X$ to $U$ measures the mutiplicative gain, upon observing $U$, of the
probability of correctly guessing a randomized function of
$X$, maximized over all such randomized functions. This is shown in \cite[Theorem 1]{Issa} to be equivalent to 
\begin{equation}\label{maxl}
\mathcal{L}(X\to U)=\log\sum_{u\in\mathcal{U}}\max_{\substack{x\in\mathcal{X}\\p_X(x)>0}}p_{U|X}(u|x).
\end{equation}
In our definition of $\mathcal{X}$, at the begining of this chapter, we assumed that $p_X(x)>0, \forall x\in\mathcal{X}$. Hence, the condition $p_X(x)>0$ can be dropped in the above definition.

The \textit{maximum information leakage}, defined in \cite{Calmon1} as
\begin{equation}
I^*(X;U)\triangleq H(X)-\min_{u\in\mathcal{U}}H(X|U=u),
\end{equation}
measures the worst-case information leakage over all the realizations of the released variable $U$.

The \textit{maximal $\alpha$-leakage} ($\mathcal{L}_{\alpha}^{max}(X\to U)$) is proposed in \cite{Liao} as a tunable measure for information leakage. The tuning parameter $\alpha$ ranges from one to infinity, where at the extremes of $\alpha=1$ and $\alpha=\infty$, it boils down to mutual information and maximal leakage, respectively.}
{\color{black}
\section{The optimal utility-privacy trade-off}\label{solv}
In this section, we address the optimal utility-privacy trade-off problem when privacy is measured by the average total variation distance given in (\ref{pr}). We consider three different utility measures, in particular, the mutual information, MMSE, and error probability. The corresponding utility-privacy trade-offs are defined as follows:
\begin{align}
m_{\epsilon}(X,Y)&\triangleq\max_{\substack{p_{U|Y}:\\X-Y-U\\T(X;U)\leq\epsilon}}I(Y;U),\label{MI}\\
M_{\epsilon}(X,Y)&\triangleq\min_{\substack{p_{U|Y}:\\X-Y-U\\T(X;U)\leq\epsilon}}\mathds{E}[(Y-U)^2],\label{MMSE}\\
E_{\epsilon}(X,Y)&\triangleq\min_{\substack{p_{U|Y}:\\X-Y-U\\T(X;U)\leq\epsilon}}\!\!\mbox{Pr}\{Y\neq U\}.\label{errorp}
\end{align}
In the following theorem, we present the optimal utility-privacy trade-off $m_\epsilon(X,Y)$ for the special case of binary $Y$, since i) it admits a closed-form solution, and ii) it can be generalized to arbitrary finite $\mathcal{Y}$.

\textbf{Theorem 1.} Let $(X,Y)\in\mathcal{X}\times\{y_1,y_2\}$ ($|\mathcal{X}|<\infty$) with $p_Y(y_1)=p$ and $\mathbf{P}_{X|Y}=\begin{bmatrix}
\mathbf{p}_{X|y_1}&\mathbf{p}_{X|y_2}
\end{bmatrix}_{|\mathcal{X}|\times2}$. We have
\begin{equation}\label{m1}
m_{\epsilon}(X,Y)=\min\left\{1,\frac{\epsilon}{p(1-p)\|\mathbf{p}_{X|y_1}-\mathbf{p}_{X|y_2}\|_1}\right\}H_b(p),
\end{equation}
\begin{proof}
Let $p_{Y|U}(y_1|u)$ be denoted by $q_u, \forall u\in\mathcal{U}$. We have 
\begin{align*}
2T(X;U)&=\sum_up_U(u)\|\mathbf{p}_{X|u}-\mathbf{p}_X\|_1\\&=\sum_up_U(u)\bigg\|\mathbf{P}_{X|Y}\left(\begin{bmatrix}
q_u\\1-q_u
\end{bmatrix}-\begin{bmatrix}
p\\1-p
\end{bmatrix}\right)\bigg\|_1\nonumber\\
&=\|\mathbf{p}_{X|y_1}-\mathbf{p}_{X|y_2}\|_1\sum_up_U(u)|q_u-p|.
\end{align*}
From the constraint $T(X;U)\leq\epsilon$, we obtain
$\sum_up_U(u)|q_u-p|\leq\frac{2\epsilon}{\|\mathbf{p}_{X|y_1}-\mathbf{p}_{X|y_2}\|_1}$.
Denoting the right hand side (RHS) of the above by $\eta$, $m_\epsilon(X,Y)$ is given by
\begin{align}
m_\epsilon(X,Y)&=\max_{\substack{p_{U|Y}:\\X-Y-U\\T(X;U)\leq\epsilon}}I(Y;U)\nonumber\\
&=H(Y)-\min_{\substack{p_{U|Y}:\\X-Y-U\\T(X;U)\leq\epsilon}}H(Y|U)\label{mc2}\\
&=H_b(p)-\!\!\!\!\!\!\!\!\min_{\substack{p_U(\cdot),q_u:\\\sum_up_U(u)|q_u-p|\leq\eta,\\ \ \sum_u p_U(u)q_u=p}}\!\!\!\!\sum_{u}p_U(u)H_b(q_u),\label{m2}
\end{align}
where the equality in (\ref{m2}) follows from the fact the constraints of minimization in (\ref{mc2}) and (\ref{m2}) are equivalent.


In what follows, we show that in the minimization in (\ref{m2}), there is no loss of optimality if instead of $q_u\in[0,1]$, we replace $q_u\in\{0,p,1\}$.

Assume that an arbitrary $U$ that satisfies $X-Y-U$ is given with its corresponding values $q_u\in[0,1],\forall u\in\mathcal{U}$, that satisfies the constraints of optimization, i.e., $\sum_up_U(u)|q_u-p|\leq\eta$ and $\sum_u p_U(u)q_u=p$.
Assume that there exists\footnote{if not, there is nothing to prove.} $u_0\in\mathcal{U}$, such that $q_{u_0}\not\in\{0,p,1\}$. Therefore, we have $q_{u_0}\in(0,p)$ or $q_{u_0}\in(p,1).$ In any case, $q_{u_0}$ can be written as a convex combination of the extreme points of the segment it belongs to. Assume that $q_{u_0}\in(p,1)$. Hence, we can write $q_{u_0}=\frac{1-q_{u_0}}{1-p}\times p+\frac{q_{u_0}-p}{1-p}\times1$. Construct the Markov chain $X-Y-U'$ as follows. Let $\mathcal{U'}\triangleq(\mathcal{U}\backslash \{u_0\})\cup\{\hat{u_0},\tilde{u_0}\}$. Let, $p_{U'}(u)=p_U(u), \forall u\in\mathcal{U}\backslash\{u_0\}, p_{U'}(\hat{u_0})=\frac{1-q_{u_0}}{1-p}p_U(u_0)$, and $p_{U'}(\tilde{u_0})=\frac{q_{u_0}-p}{1-p}p_U(u_0)$. Finally, let $q_{u'}$ remain unchanged for all the elements of $\mathcal{U}\backslash\{u_0\}$, and $q_{\hat{u_0}}=p, q_{\tilde{u_0}}=1$. Due to linearity, it can be readily verified that $\sum_{u'\in\mathcal{U'}}p_{U'}(u')|q_{u'}-p|=\sum_up_U(u)|q_u-p|\leq\eta$ and $\sum_{u'\in\mathcal{U'}} p_{U'}(u')q_{u'}=\sum_u p_U(u)q_u=p$. Hence, $p_{U'|Y}$ is in the feasible region of the optimization. Furthermore, from the concavity of entropy, we have
\begin{align*}
\sum_{u\in\mathcal{U}}p_U(u)H_b(q_u)&=\sum_{u\in\mathcal{U}\backslash\{u_0\}}p_U(u)H_b(q_u)+p_U(u_0)H_b(q_{u_0})\\
&=\sum_{u\in\mathcal{U}\backslash\{u_0\}}p_U(u)H_b(q_u)\\
&\ \ \ +p_U(u_0)H_b(\frac{1-q_{u_0}}{1-p}\times p+\frac{q_{u_0}-p}{1-p}\times1)
\end{align*}
\begin{align*}
&\geq \sum_{u\in\mathcal{U}\backslash\{u_0\}}p_U(u)H_b(q_u)+p_{U'}(\hat{u_0})H_b(q_{\hat{u_0}})\\
&\ \ \ +p_{U'}(\tilde{u_0})H_b(q_{\tilde{u_0}})\\
&=\sum_{u'\in\mathcal{U'}} p_{U'}(u')H_b(q_{u'})
\end{align*}
Therefore, the performance of the privacy mapping $p_{U'|Y}$ is at least as good as\footnote{Note that entropy is strictly concave, and $p_{U'|Y}$ outperforms $p_{U|Y}$. Nevertheless, what is sufficient in this analysis is just concavity.} that of $p_{U|Y}$. Therefore, without loss of optimality, the constraint $q_u\in(1,p)$ can be replaced by $q_u\in\{p,1\}$. In a similar way, $q_u\in(0,p)$ can be replaced by $q_u\in\{0,p\}$, which results in the sufficiency of $q_u\in\{0,p,1\}$.
Therefore, setting $\mathcal{U}=\{u_1,u_2,u_3\}$ in direct correspondence to $\{0,p,1\}$, the problem resuces to the following linear program
\begin{align*}
\max_{\substack{p_U(\cdot):\\p_U(u_1)p+p_U(u_3)(1-p)\leq\eta\\p_U(u_2)p+p_U(u_3)=p}}\!\!\!\!\!\!\!\!\! \left(1-p_U(u_2)\right)H_b(p),
\end{align*}
which can be readily found to be equal to $\min\left\{1,\frac{\eta}{2p(1-p)}\right\}\cdot H_b(p).$
Replacing $\eta$ with $\frac{2\epsilon}{\|\mathbf{p}_{X|y_1}-\mathbf{p}_{X|y_2}\|_1}$ results in (\ref{m1}).
\end{proof}

\textbf{Remark 1.} The proof of Theorem 1 relies on a simple fact: the minimum of a concave function over a convex set is attained at an extreme point of that set. 

The following theorem, whose proof is provided in Appendix \ref{app0}, generalizes Theorem 1 and relies on the concavity/convexity of the objective function and piece-wise linearity of the $L^1$-norm.

}
\textbf{Theorem 2.} For a pair of random variables $(X,Y)\in\mathcal{X}\times\mathcal{Y}$ ($|\mathcal{X}|,|\mathcal{Y}|<\infty$), $m_{\epsilon}(X,Y), M_{\epsilon}(X,Y)$ and $E_{\epsilon}(X,Y)$ are the solutions to a standard linear program (LP).

\section{Motivation of total variation distance as a measure of privacy}\label{jus}
The following three subsections motivate the use of total variation distance as a measure of privacy.
\subsection{Post-processing and linkage inequalities}
For an arbitrary privacy-leakage measure $J(X;U)$, we have the following definitions from \cite{Ishwar}.

\textbf{Definition 1. (\textit{Post-processing inequality})} $J$ satisfies the post-processing inequality if and only if for any Markov chain $A-B-C$, we have $J(A;B)\geq J(A;C)$.

\textbf{Definition 2. (\textit{Linkage inequality})} $J$ satisfies the linkage inequality if and only if for any Markov chain $A-B-C$, we have $J(B;C)\geq J(A;C)$.

It is obvious that for a symmetric privacy measure, i.e., $J(X;U)=J(U;X)$, like mutual information, the two definitions are equivalent. As mentioned in \cite{Ishwar}, the post-processing inequality captures an intuitive axiomatic requirement that no independent post-processing of the data can increase the privacy-leakage. On the other hand, the linkage inequality states that if we have primary and secondary sensitive data ($B$ and $A$, respectively), and the released data $C$ is generated independently from only the primary sensitive data, then the privacy-leakage of the secondary data is bounded by that of the primary data. As an additional note, it is shown in \cite{Ishwar} that not all of the privacy measures satisfy the linkage inequality, e.g., \textit{differential privacy} or \textit{maximal information leakage}\footnote{{\color{black} One of the advantages of satisfying the linkage inequality is as follows. Consider the same scenario $X-Y-U$, where the distribution of the private data $X$ is unknown, or complex to learn. If we can find $X'$ satisfying $X-X'-Y-U$ whose distribution is known or at least easily learnable, then satisfying the linkage inequality is beneficial in the sense that by keeping the privacy of $X'$, privacy of $X$ is preserved, i.e., $J(X;U)\leq J(X';U)\leq \epsilon.$ This is simply a case of having layers of private information. Also, consider the case where the privacy of any private latent variable $X$ that satisfies $X-Y-U$ should be preserved by the release mechanism. Then, if linkage inequality is satisfied, the solution would simply be $J(Y;U)\leq\epsilon$.}}.

\textbf{Theorem 3.} The privacy measure $T(\cdot;\cdot)$ given in (\ref{pr}) satisfies both the post-processing and the linkage inequalities.
\begin{proof}
Let $A-B-C$ form a Markov chain. We have
\begin{align}
2T(A;B)&=\sum_{b}p_B(b)\|\mathbf{p}_{A|b}-\mathbf{p}_A\|_1\nonumber\\&=\sum_{b,c}p_{B,C}(b,c)\|\mathbf{p}_{A|b,c}-\mathbf{p}_A\|_1\label{kh1}\\
&=\sum_{c}p_C(c)\sum_bp_{B|C}(b|c)\|\mathbf{p}_{A|b,c}-\mathbf{p}_A\|_1\nonumber\\
&\geq\sum_{c}p_C(c)\bigg\|\sum_bp_{B|C}(b|c)\mathbf{p}_{A|b,c}-\mathbf{p}_A\bigg\|_1\label{kh2}\\
&=\sum_{c}p_C(c)\|\mathbf{p}_{A|c}-\mathbf{p}_A\|_1\nonumber\\
&=2T(A;C),
\end{align}
where (\ref{kh1}) follows from the fact that $A-B-C$ form a Markov chain; (\ref{kh2}) results from the convexity of the $L^1$-norm. This proves the post-processing inequality.

In order to prove that $T(\cdot;\cdot)$, given in (\ref{pr}), satisfies the linkage inequality, we can write
\begin{align}
2T(A;C)&=\sum_{c}p_C(c)\|\mathbf{p}_{A|c}-\mathbf{p}_A\|_1\nonumber\\
&=\sum_{c}p_C(c)\|\mathbf{P}_{A|B}(\mathbf{p}_{B|c}-\mathbf{p}_B)\|_1\nonumber\\
&=\sum_{c}p_C(c)\sum_a\bigg|\sum_bp_{A|B}(a|b)\left(p_{B|C}(b|c)-p_B(b)\right)\!\!\bigg|\nonumber
\end{align}
\begin{align}
&\leq\sum_{c}p_C(c)\sum_a\sum_bp_{A|B}(a|b)|p_{B|C}(b|c)-p_B(b)|\label{kh3}\\
&=\sum_{c}p_C(c)\sum_b\sum_ap_{A|B}(a|b)|p_{B|C}(b|c)-p_B(b)|\nonumber\\
&=\sum_{c}p_C(c)\|\mathbf{p}_{B|c}-\mathbf{p}_B\|_1\nonumber\\
&=2T(B;C),
\end{align}
where (\ref{kh3}) follows from the triangle inequality. 
\end{proof}

\textbf{Remark 2.} Among all the $L^p$-norms ($p\geq 1$), only the $L^1$-norm satisfies the linkage inequality. Consider the following example: Let $A-B-C$ form a Markov chain, and consider the transition matrix
\begin{equation*}
\mathbf{P}_{A|B}=\begin{bmatrix}
1&1&\frac{1}{2}&0&0&0\\
0&0&\frac{1}{2}&1&1&0\\
0&0&0&0&0&1
\end{bmatrix},
\end{equation*}
\begin{figure}[t]
  \centering
  \includegraphics[scale=0.5]{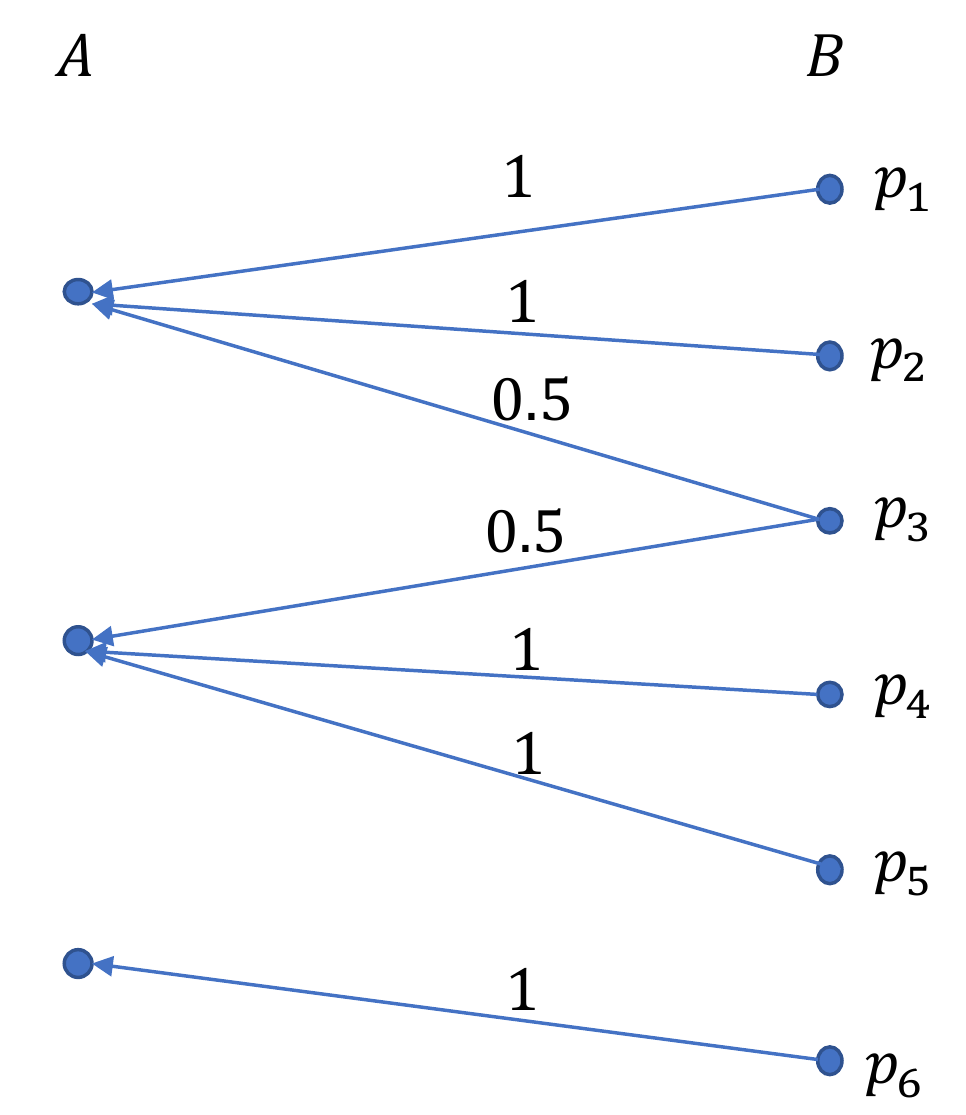}\\
  \caption{The example in Remark 2.}\label{fig2}
\end{figure}
as shown in Figure \ref{fig2} with $\mathbf{p}_B=\begin{bmatrix}
p_1&p_2&p_3&p_4&p_5&p_6
\end{bmatrix}^T$, where $p_i\in(0,1),\forall i\in[1:6]$ and $\sum_{i=1}^6p_i=1$. Let $C\in\{-1,1\}$, and $p_C(1)=\frac{1}{2}$. For sufficiently small $\delta>0$, let $\mathbf{p}_{B|c}=\begin{bmatrix}
p_1+c\delta&p_2+c\delta&p_3&p_4-c\delta&p_5-c\delta&p_6
\end{bmatrix}^T$ which results in $\mathbf{p}_{A|c}=\begin{bmatrix}
p_7+2c\delta&p_8-2c\delta&p_6
\end{bmatrix}^T,\forall c\in\{-1,1\}$. It can be verified that for any $p\in(1,+\infty]$, we have $\|\mathbf{p}_{A|c}-\mathbf{p}_A\|_p>\|\mathbf{p}_{B|c}-\mathbf{p}_B\|_p, \forall c\in\{-1,1\}$.

Note that the quantity $\|\mathbf{x}\|_p=(\sum_i|x_i|^p)^{\frac{1}{p}}$ is not subadditive when $p\in(0,1)$, and thus, does not define a norm. Nonetheless, even if the privacy measure is defined as $J(A;B)=\sum_bp_B(b)\|\mathbf{p}_{A|b}-\mathbf{p}_A\|_p$ with $p\in(0,1)$, it can be verified that it does not satisfy the linkage inequality by letting $\mathbf{p}_{B|c}=\begin{bmatrix}
p_1&p_2&p_3+c\delta&p_4&p_5&p_6-c\delta
\end{bmatrix}^T,\forall c\in\{-1,1\}$, in the counterexample of this remark.

\textbf{Remark 3.} It is obvious that from the post-processing inequality, the feasible range of $\epsilon$ can be tightened to $[0,T(X;Y)]$.
\subsection{Bounding inference threats}
An inference threat model is introduced in \cite{Calmon1}, which models a broad class of statistical inference attacks that can be performed on private data $X$. Assume that an inference cost function $C(\cdot,\cdot):\mathcal{X}\times\mathcal{P}(\mathcal{X})\to\mathbb{R}$ is given. Prior to observing $U$, the attacker chooses a belief distribution $\mathbf{q}$ over $X$ as the solution of $c_0^*=\min_{\mathbf{q}\in\mathcal{P}(\mathcal{X})}\mathds{E}_X[C(X,\mathbf{q})]$, where the minimizer is denoted by $\mathbf{q}_0^*$, while after observing $U=u$, he revises this belief as the solution of $c_u^*=\min_{\mathbf{q}\in\mathcal{P}(\mathcal{X})}\mathds{E}_{X|U}[C(X,\mathbf{q})|U=u]$, where the minimizer is denoted by $\mathbf{q}_u^*$. As a result, the attacker obtains an average gain in inference cost of $\Delta C=c_0^*-\mathds{E}_U[c_U^*]$, which quantifies the improvement in his inference. A natural way to restrict the attacker's inference quality is to keep $\Delta C$ below a target value. The following theorem ensures that for any bounded cost function $C(\cdot,\cdot)$, the attacker's inference quality is restricted in this way by focusing on the control of $T(X;U)$, i.e., keeping it below a certain threshold.

\textbf{Theorem 4.} Let $L=\sup_{x\in\mathcal{X},\mathbf{q}\in\mathcal{P}(\mathcal{X})}|C(x,\mathbf{q})|<+\infty$. We have $\Delta C\leq 4L\cdot T(X;U)$.
\begin{proof}
The proof follows similar steps as in \cite[Lemma 2]{Makhdoumi} up to the point of using Pinsker inequality, which is restated here.
\begin{align}
\Delta C&=c_0^*-\mathds{E}_U[c_U^*]\nonumber\\
& = \mathds{E}_X[C(X,\mathbf{q}_0^*)]-\mathds{E}_U\bigg[\mathds{E}_{X|U}[C(X,\mathbf{q}_U^*)|U=u]\bigg]\nonumber\\
&=\mathds{E}_U\bigg[\mathds{E}_{X|U}[C(X,\mathbf{q}_0^*)-C(X,\mathbf{q}_U^*)|U=u]\bigg]\nonumber\\
&=\mathds{E}_U\bigg[\mathds{E}_{X|U}[C(X,\mathbf{q}_0^*)-C(X,\mathbf{q}_U^*)|U=u]\nonumber\\&\ \ \ -\mathds{E}_X[C(X,\mathbf{q}_0^*)-C(X,\mathbf{q}_U^*)]\nonumber\\
&\ \ \ +\mathds{E}_X[C(X,\mathbf{q}_0^*)-C(X,\mathbf{q}_U^*)]\bigg]\nonumber\\
&\leq\mathds{E}_U\bigg[\sum_x(p_{X|U}(x|U)-p_X(x))(C(x,\mathbf{q}_0^*)-C(x,\mathbf{q}_U^*))\bigg]\label{esb}\\
&\leq\mathds{E}_U\bigg[2L\sum_x|p_{X|U}(x|U)-p_X(x)|\bigg]\label{esb2}\\
&= 4L\cdot T(X;U),
\end{align}
where (\ref{esb}) follows from the fact that $\mathbf{q}_0^*$ is the minimizer of $\mathds{E}_X[C(X,\mathbf{q})]$ over $\mathcal{P}(\mathcal{X})$, and therefore, $\mathds{E}_X[C(X,\mathbf{q}_0^*)-C(X,\mathbf{q}_u^*)]\leq 0$; in (\ref{esb2}), the assumption $|C(\cdot,\cdot)|\leq L$ has been used.
\end{proof}
\color{black}
In the following theorem, it is shown that the privacy measure proposed in this paper, i.e., $T(X;U)$, can serve as lower and upper bounds for mutual information and maximal leakage.

\textbf{Theorem 5.} The following upper and lower bounds hold.
\begin{align}
I(X;U)&\geq 2\log_2e\cdot T^2(X;U)\label{bound1}\\
\mathcal{L}(X\to U)&\leq\log\bigg(1+\frac{T(X;U)}{\min_xp_X(x)}\bigg)\label{bound2}\\
\mathcal{L}(X\to U)&\geq\log\bigg(1+\frac{T(X;U)}{(|\mathcal{X}|-1)\max_xp_X(x)}\bigg)\label{bound3}
\end{align}
The proof of this Theorem is provided in Appendix \ref{app1.5}.

\textbf{Remark 4.} It is known from \cite{Liao} that $ I(X;U)
\leq\mathcal{L}_{\alpha}^{max}(X\to U)
\leq \mathcal{L}(X\to U)$. Therefore, combined with the bounds in Theorem 5, we can write
\begin{align}
2\log_2e\cdot T^2(X;U)&\leq I(X;U)\nonumber\\
&\leq \mathcal{L}_{\alpha}^{max}(X\to U)\nonumber\\
&\leq \mathcal{L}(X\to U)\nonumber\\
&\leq \log\bigg(1+\frac{T(X;U)}{\min_xp_X(x)}\bigg)\label{superbound}.
\end{align}
\textbf{Remark 5.} It is important to note that, in bounding the inference gain of an adversary by $T(X;U)$ (as in the beginning of this subsection), the boundedness of the cost function is not a necessary condition. For example, the log-loss cost function, i.e., $C(x,\mathbf{q})=-\log q(x)$, where $q(\cdot)$ is the pmf corresponding to $\mathbf{q}$, is not a bounded cost function. However, $\Delta C$ under log-loss cost function, which is equal to $I(X;U)$, is bounded above by $T(X;U)$ as in (\ref{superbound}).

\begin{figure}[t]
  \centering
  \includegraphics[scale=0.25]{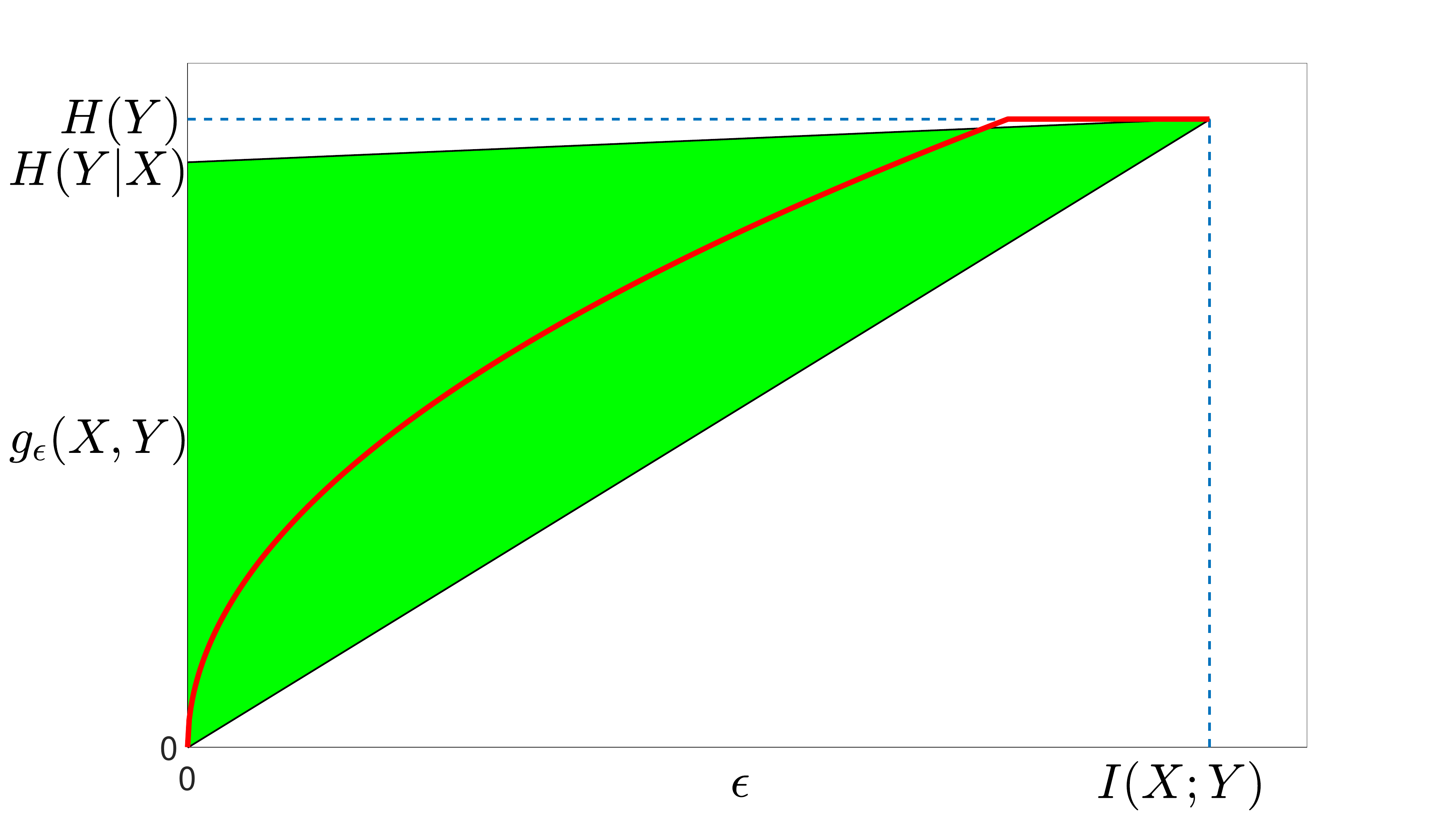}\\
  \caption{Tightening the permissible region of the utility-privacy trade-off by employing $T(X;U)$.}\label{fig66}
\end{figure}

\textbf{Remark 6.}
It is interesting to note that as a by-product of the lower bound in (\ref{bound1}) and Theorem 1, we can get non-trivial bounds for the following quantity
\begin{equation*}
g_\epsilon(X,Y)=\max_{\substack{U:X-Y-U\\I(X;U)\leq \epsilon}}I(Y;U),
\end{equation*}
which is the utility-privacy trade-off when mutual information is employed as both the utility and privacy measure \cite{Shahab1}.
From \cite[Lemma 1]{SA2}, we have 
\begin{equation}\label{shahabi}
\frac{H(Y)}{I(X;Y)}\epsilon\leq g_\epsilon(X,Y)\leq \epsilon+H(Y|X),\ \ \  \epsilon\in[0,I(X;Y)].
\end{equation}
The upper and lower bounds are two lines shown in Figure \ref{fig66}. 
Assume that $Y$ is binary with $p_Y(y_1)=p$, and $X$ is an arbitrary discrete random variable.
Assume that instead of $I(X;U)$, we use its lower bound in Theorem 5, i.e., $2\log_2e\cdot T^2(X;U)$. Hence, by weakening the constraint, we have an upper bound for the objective function as
\begin{align}\label{nont}
g_\epsilon(X,Y)&\leq \max_{\substack{U:X-Y-U\\T(X;U)\leq \sqrt{\frac{\epsilon}{2\log_2e}}}}I(Y;U)\nonumber\\
&=\min\left\{1,\frac{\sqrt{\frac{\epsilon}{2\log_2e}}}{p(1-p)\|\mathbf{p}_{X|y_1}-\mathbf{p}_{X|y_2}\|_1}\right\}H_b(p),
\end{align}
which follows from Theorem 1.
Figure \ref{fig66} shows the upper bound of (\ref{nont}), along with the two straight lines denoting the upper and lower bounds in (\ref{shahabi}) for the following example:
 $(X,Y)\in\mathcal{X}\times\mathcal{Y}$, where $\mathcal{X}=\{x_1,x_2,x_3\}$ and $\mathcal{Y}=\{y_1,y_2\}$.
\begin{equation*}
\mathbf{P}_{X|Y}=\begin{bmatrix}
0.5&0.3\\0.3&0.2\\0.2&0.5
\end{bmatrix},\ \mathbf{p}_Y=\begin{bmatrix}
\frac{1}{3}\\ \frac{2}{3}
\end{bmatrix}
\end{equation*}
As it can be seen, this is a non-trivial bound that has further tightened the permissible region for the utility-privacy trade-off.

\color{black}
\subsection{Evaluation of the optimal utility-privacy trade-off}
As shown in this paper, the optimal utility-privacy trade-offs in (\ref{MI}) to (\ref{errorp}) reduce to an LP when $T(X;U)$ is employed as the privacy measure. This result follows from the concavity of the objective functions and piece-wise linearity of the $L_1$-norm\footnote{This is also the case in the more general observation model in \cite{Ishwar}, i.e., for the Markov chain $(X,Y)-W-U$.}. Examples of these trade-off regions are provided in Section \ref{nmu} for different utility measures. Other measures of privacy do not necessarily lend themselves to exact characterization. For example, when mutual information is considered as both the privacy and utility measures, the characterization of the optimal trade-off ($g_\epsilon(X;Y)$ in \cite{Shahab1}) is an open problem. {\color{black} Another example is the trade-off when $\chi^2$-based information measures capture both utility and privacy, for which upper and lower bounds are proposed in \cite{Wang2}, and for a special case, a convex program is developed to solve the trade-off. The fact that the exact utility-privacy trade-off under $T(X;U)$ can be solved is not only important on its own, but also benefical in bounding the trade-offs under other privacy measures, as mentioned in Remark 6.}

{\color{black}
\textbf{Remark 7.}
We emphasize here that the analysis in this paper relies on the fact that the joint distribution of the private and available data are known and can be fed as an input to the release mechanism, as in \cite{Rebollo} and\cite{Calmon1}. In practice, the true data distribution may not always be available, and therefore, further analysis based on learning methods is needed to address the utility-privacy trade-off. In this regard, \cite{Huang1, Ish22, chenchen} propose a training method based on the application of Generative Adversarial Networks (GAN) framework \cite{NIPS}, which can be captured as a minimax game between two parties, as a data-driven approach to address this problem. As a related work, \cite{Wang3} analyzes the performance of privacy-preserving release mechanisms under partial knowledge of the input distribution for different privacy measures. It is important to note that the proposed privacy measure, i.e., $T(\cdot;\cdot)$ guarantees pointwise and uniform privacy according to \cite[Theorems 1,2]{Wang3}. An extension of the current work is to address the utility-privacy trade-off under the privacy measure $T(X;U)$ when only a limited number of observed data samples are available to the release mechanism.

\textbf{Remark 8.} It is interesting to note that full knowledge of the joint distribution $p_{X,Y}$, is not necessary for the privacy-preserving release mechanism under our proposed privacy measure $T(X;U)$. For instance, according to Theorem 1, the privacy-preserving release mechanism has to know the joint distribution $\mathbf{P}_{X,Y}$ only through two quantities $p_Y(y_1)$ and $\|\mathbf{p}_{X|y_1}-\mathbf{p}_{X|y_2}\|_1$, rather than $2|\mathcal{X}|-1$ quantities that fully capture the joint distribution. In this regard, another interesting problem is to evaluate the minimum amount of information that is needed by the release mechanism.
}
\\

\section{Numerical results}\label{nmu}
{\color{black}
Here, we provide some numerical examples for the optimal utility-privacy trade-off under total variation distance as the privacy measure. Assume that the pair $(X,Y)$ is distributed according to the joint distribution given in Figure \ref{fig3}.
\begin{figure}[t]
  \centering
  \includegraphics[scale=0.5]{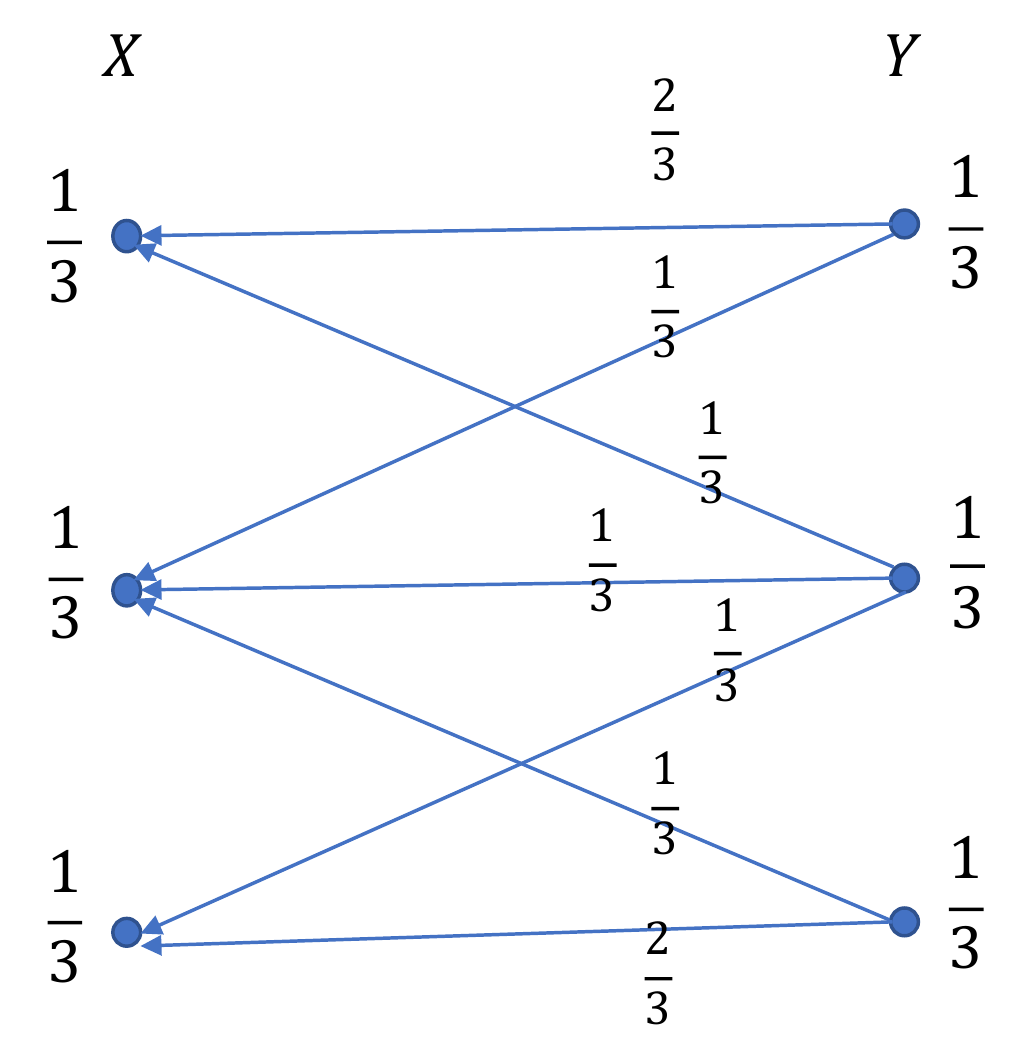}\\
  \caption{An example joint distribution $p_{X,Y}$, where $\mathbf{p}_X=\mathbf{p}_Y=[\frac{1}{3}\ \frac{1}{3}\ \frac{1}{3}]^T$, and $p_{X|Y}$ is according to the figure.}\label{fig3}
\end{figure}
Figure \ref{fig6} captures the trade-offs in (\ref{MI}) to (\ref{errorp}). In the evaluation of $M_{\epsilon}(X,Y)$, we have assumed $\mathcal{Y}=\{y_1,y_2,y_3\}=\{1,0,-1\}$.

In the evaluation of the utility-privacy trade-off, the LP can be solved by simplex method, which has polynomial-time average-case complexity, however, as it can be observed in the proof of Theorem 2, we need to check at most $2^{|\mathcal{X}|}$ regions in $\mathcal{P}(\mathcal{Y})$ based on the sign of $|\mathcal{X}|$ elements of the $L^1$-norm, which grows exponentially with $|\mathcal{X}|$. However, it is important to note that this is the worst case, as for example, no matter how large $|\mathcal{X}|$ is, we have only two regions when $Y$ is binary.
}


\begin{figure}[t]
  \centering
  \includegraphics[scale=0.3]{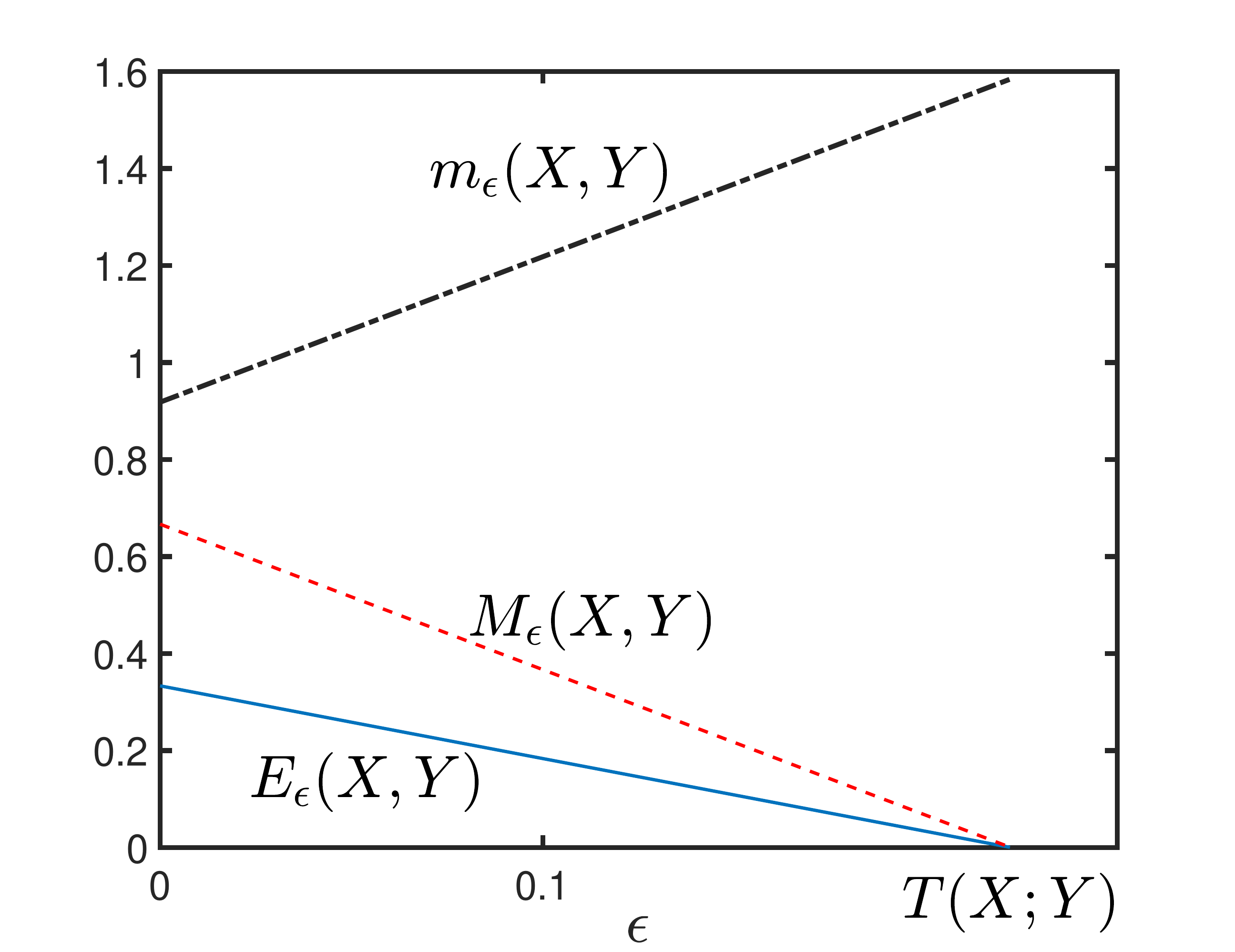}\\
  \caption{The optimal utility-privacy trade-off regions.}\label{fig6}
\end{figure}
\section{Conclusions}
We have introduced and motivated total variation distance as an information-theoretic privacy-leakage measure by showing that i) it satisfies the \textit{post-processing} and \textit{linkage} inequalities; ii) the corresponding optimal utility-privacy trade-off can be solved through a standard linear program; and iii) it provides a bound on the privacy-leakage measured by the mutual information, the \textit{maximal leakage}, or the improvement in an inference attack with a bounded cost function.
\appendices
{\color{black}
\section{Proof of theorem 2}\label{app0}
Let $\psi(\cdot)$ be a continuous and concave functional defined on $\mathcal{P}(\mathcal{Y}).$ The following Proposition serves a the main part of this proof.

\textbf{Proposition 1.} In the following optimization problem 
\begin{equation}\label{minop}
\min_{\substack{F_U(\cdot),\ \mathbf{p}_{Y|u}\in\mathcal{P}(\mathcal{Y}):\\\frac{1}{2}\int_{\mathcal{U}}\|\mathbf{p}_{X|u}-\mathbf{p}_X\|_1dF(u)\leq\epsilon \\ \int_{\mathcal{U}} \mathbf{p}_{Y|u}dF(u)=\mathbf{p}_Y}}\int\psi(\mathbf{p}_{Y|u})dF_U(u),
\end{equation}
it is sufficient to have $|\mathcal{U}|\leq|\mathcal{Y}|+1$, and the solution is obtained by a linear program.
\begin{proof}
For all $ \mathbf{x}\in\mathcal{P}(\mathcal{Y})$, consider the following quantity
\begin{equation}\label{efraz}
f(\mathbf{x})\triangleq\|\mathbf{P}_{X|Y}(\mathbf{x}-\mathbf{p}_Y)\|_1=\sum_{i=1}^{|\mathcal{X}|}|\mathbf{r}_i(\mathbf{x}-\mathbf{p}_Y)|,
\end{equation}
where $\mathbf{r}_i$ denotes the $i$-th row of matrix $\mathbf{P}_{X|Y}$. Based on where $\mathbf{x}$ is located on $\mathcal{P}(\mathcal{Y})$, each argument in the absolute value in (\ref{efraz}), can be negative or non-negative. Hence, the quantity in (\ref{efraz}) divides $\mathcal{P}(\mathcal{Y})$ into at most $2^{|\mathcal{X}|}$ partitions, i.e., $\mathcal{P}(\mathcal{Y})=\cup_{i=1}^{K}\mathbb{S}_i$, where $K\leq2^{|\mathcal{X}|}$. It can be readily verified that each $\mathbb{S}_i$ is a convex polytope with a finite number of extreme points ( since it can be written as the intersection of a finite number of closed half-spaces in $\mathcal{P}(\mathcal{Y})$) and for $x\in\mathbb{S}_i$, $f(\mathbf{x})$ is linear in $\mathbf{x}$. Let $\hat{\mathbb{S}_i}$ denote the set of extreme points of $\mathbb{S}_i$, and $\mathbb{S}\triangleq\cup_{i=1}^K\hat{\mathbb{S}_i}$. In the minimization in (\ref{minop}), it is sufficient to replace $\mathbf{p}_{Y|u}\in\mathcal{P}(\mathcal{Y})$ with $\mathbf{p}_{Y|u}\in\mathbb{S}$. This is simply a generalization of the proof of Theorem 1, and relies on the concavity of $\psi(\cdot)$ and linearity of $f(\cdot)$ over any $\mathbb{S}_i$. In other words, any $\mathbf{p}_{Y|u}$ can be written as a convex combination of the extreme points of the set it belongs to (i.e., $\mathbb{S}_i$ for some $i\in[1:K]$), while preserving the constraint of optimization and not increasing the objective function. When the objective function is strictly concave, this procedure decreases the objective function.

Once the elements of $\mathbb{S}=\{\mathbf{s}_1,\mathbf{s}_2,\ldots,\mathbf{s}_K\}$ are identified, the problem in (\ref{minop}) reduces to
\begin{equation}\label{minop2}
\min_{\substack{p_U(\cdot):\\\frac{1}{2}\sum_{i=1}^Kp_U(u_i)\|\mathbf{P}_{X|Y}(\mathbf{s}_i-\mathbf{p}_X)\|_1\leq\epsilon \\\sum_{i=1}^Kp_U(u_i)\mathbf{s}_{i}=\mathbf{p}_Y}}\sum_{i=1}^Kp_U(u_i)\psi(\mathbf{s}_{i}),
\end{equation}
which is a linear program. It can be verified that the constraint $\sum_{i=1}^Kp_U(u_i)=1$ is satisfied if the second constraint in the LP, i.e., $\sum_{i=1}^Kp_U(u_i)\mathbf{s}_{i}=\mathbf{p}_Y$ is met. Finally, the procedure of finding the elements of $\mathbb{S}$ is provided in Appendix \ref{app2}.

Showing $|\mathcal{U}|\leq|\mathcal{Y}|+1$ follows the routine application of cardinality bounding techniques (e.g. \cite{SA2}) as follows. Let $\mathbf c :\mathcal{P}(\mathcal{Y})\to\mathbb{R}^{|\mathcal{Y}|+1}$ be a vector-valued mapping defined element-wise as
\begin{align*}
c_i(p_{Y|U}(\cdot|u))&=p_{Y|U}(y_i|u),\ i\in[1:|\mathcal{Y}|-1]\\
c_{|\mathcal{Y}|}(p_{Y|U}(\cdot|u))&=\psi(\mathbf{p}_{Y|u}),\\
c_{|\mathcal{Y}|+1}(p_{Y|U}(\cdot|u))&=\frac{1}{2}\|\mathbf{P}_{X|Y}(\mathbf{p}_{Y|u}-\mathbf{p}_Y)\|_1
\end{align*}
Since $\mathcal{P}(\mathcal{Y})$ is a closed and bounded subset of $\mathbb{R}^{|\mathcal{Y}|}$, it is compact. Also, $\mathbf{c}$ is a continuous mapping. Therefore, from the support lemma \cite{Elgamal}, for every $U\sim F(u)$ defined on (arbitrary) $\mathcal{U}$, there exists a random variable $U'\sim p(u')$ with $|\mathcal{U'}|\leq |\mathcal{Y}|$ and a collection of conditional pmfs $p_{Y|U'}(\cdot|u')$ indexed by $u'\in\mathcal{U}'$, such that
\begin{equation*}
\int_{\mathcal{U}}c_i(p(y|u))dF(u)=\sum_{u'\in\mathcal{U'}}c_i(p(y|u'))p(u'),\ i\in[1:|\mathcal{Y}|].
\end{equation*}
Therefore, there is no loss of optimality in considering $|\mathcal{U}|\leq|\mathcal{Y}|+1$.
\end{proof}
}
The utility-privacy trade-off in (\ref{MI}) can be rewritten as
\begin{align}
m_\epsilon(X,Y)
&=H(Y)-\min_{\substack{p_U(\cdot),\mathbf{p}_{Y|u}:\\T(X;U)\leq\epsilon\\ \sum_{u} p_U(u)\mathbf{p}_{Y|u}=\mathbf{p}_Y}}H(Y|U)\label{min},
\end{align}
and since $H(\cdot)$ is a concave function, from (\ref{minop}), it reduces to
\begin{align}
m_{\epsilon}(X,Y)=H(Y)-\!\!\!\!\!\!\!\!\!\!\!\!\!\!\!\!\!\!\!\!\!\!\!\!\!\!\!\!\!\!\!\!\!\!\!\!\!&\min_{\substack{\mathbf{w}\geq 0:\\\begin{bmatrix}
f(\mathbf{s}_1)&f(\mathbf{s}_2)&\dots&f(\mathbf{s}_K)
\end{bmatrix}\cdot\mathbf{w}\leq2\epsilon \\\begin{bmatrix}
\mathbf{s}_1&\mathbf{s}_2&\dots&\mathbf{s}_K
\end{bmatrix}\cdot\mathbf{w}=\mathbf{p}_Y}}\ \!\!\!\!\!\!\!\!\!\!\!\!\!\!\!\!\!\!\!\!\!\!\!\!\!\!\!\!\!\!\!\!\!\!\!\!\!\begin{bmatrix}
H(\mathbf{s}_1)&H(\mathbf{s}_2)&\!\!\!\dots\!\!\!&H(\mathbf{s}_K)
\end{bmatrix}\cdot\mathbf{w}.\label{LP}
\end{align}
For the evaluation of the utility-privacy trade-off in (\ref{MMSE}), we can write
\begin{align}
\mathds{E}_{U,Y}[(Y-U)^2]&=\mathds{E}_U\bigg[\mathds{E}_{Y|U}[(Y-U)^2|U]\bigg]\nonumber\\
&\geq\mathds{E}_U\bigg[\mathds{E}_{Y|U}\left[(Y-\mathds{E}[Y|U])^2|U\right]\bigg]\label{hos1}\\
&=\int\mbox{Var}[Y|U=u]dF_U(u),\label{hos2}
\end{align}
where (\ref{hos1}) is a classical result from MMSE estimation \cite{Levy}. From (\ref{hos2}) and (\ref{MMSE}), we have the following lower bound:
\begin{equation}\label{ok}
M_{\epsilon}(X,Y)\geq\min_{\substack{F_U(\cdot),\ \mathbf{p}_{Y|u}:\\ T(X;U)\leq\epsilon\\\int_{\mathcal{U}} \mathbf{p}_{Y|u}dF(u)=\mathbf{p}_Y}}\int\mbox{Var}[Y|U=u]dF_U(u),
\end{equation}
which is tight if and only if $\mathds{E}[Y|U=u]=u,\forall u\in\mathcal{U}$. 

\textbf{Proposition 2.} $\mbox{Var}[Y|U=u]$ is a concave function of $\mathbf{p}_{Y|u}$.

The proof of this Proposition is provided in Appendix \ref{app1}.
From the concavity of $\mbox{Var}[Y|U=u]$ in Proposition 2, we can use the result of Proposition 1 and write
\begin{align}
M_{\epsilon}(X,Y)=\!\!\!\!\!\!\!\!\!\!\!\!\!\!\!\!\!\!\!\!\!\!\!\!\!\!\!\!\!\!\!&\min_{\substack{\mathbf{w}\geq 0:\\\begin{bmatrix}
f(\mathbf{s}_1)&f(\mathbf{s}_2)&\dots&f(\mathbf{s}_K)
\end{bmatrix}\cdot\mathbf{w}\leq2\epsilon \\\begin{bmatrix}
\mathbf{s}_1&\mathbf{s}_2&\dots&\mathbf{s}_K
\end{bmatrix}\cdot\mathbf{w}=\mathbf{p}_Y}}\ \!\!\!\!\!\!\!\!\!\!\!\!\!\!\!\!\!\!\!\!\!\!\!\!\!\!\!\!\!\!\!\!\!\!\begin{bmatrix}
\mbox{Var}_1&\mbox{Var}_2&\dots&\mbox{Var}_K
\end{bmatrix}\cdot\mathbf{w}
,\label{LP2}
\end{align}
where $\mbox{Var}_i$ ($\forall i\in[1:K]$) denotes $\mbox{Var}[Y|U=u]$ under $\mathbf{s}_i$, i.e., when $\mathbf{p}_{Y|u}=\mathbf{s}_i$. Finally, once the LP in (\ref{LP2}) is solved, 
if $w_i^*\neq 0$ ($i\in[1:K]$), we set $u_i=\mathds{E}[Y|U=u_i]$, where the expectation is taken over the distribution $\mathbf{p}_{Y|u_i}=\mathbf{s}_i$. 

Finally, similarly to Theorem 1, it can be verified that when $Y$ is binary, the problem in (\ref{MMSE}) has a closed form solution given by
{\color{black}
\begin{equation}
M_{\epsilon}(X,Y)=\bigg(p(1-p)-\frac{\epsilon}{\|\mathbf{p}_{X|y_1}-\mathbf{p}_{X|y_2}\|_1}\bigg)^+\cdot(y_1-y_2)^2,
\end{equation}
where $(x)^+\triangleq\max\{0,x\}$. }

For the evaluation of the utility-privacy trade-off in (\ref{errorp}), we can write
\begin{align}
\mbox{Pr}\{Y\neq U\}&=1-\mbox{Pr}\{Y= U\}\nonumber\\
&= 1- \int_{\mathcal{U}}\mbox{Pr}\{Y=u|U=u\}dF_U(u)\nonumber\\
&\geq 1 - \int_{\mathcal{U}}\max_{y}p_{Y|U}(y|u)dF_U(u),\label{asw}
\end{align}
where (\ref{asw}) holds with equality when $u = \argmax_{y}p_{Y|U}(y|u)$. Then, (\ref{errorp}) is lower bounded by
\begin{equation}\label{tok}
1+\!\!\!\!\!\min_{\substack{F_U(\cdot),\ \mathbf{p}_{Y|u}:\\T(X;U)\leq\epsilon\\ \int_{\mathcal{U}} \mathbf{p}_{Y|u}dF(u)=\mathbf{p}_Y}}\int_{\mathcal{U}}-\max_{y}p_{Y|U}(y|u)dF_U(u).
\end{equation}
For any two arbitrary pmfs $p_Y^1(\cdot)$ and $p_Y^2(\cdot)$, we have
\begin{align*}
&\max_{y}\{\alpha p_{Y}^1(y)+(1-\alpha)p_Y^2(y)\}\\
&\leq\max_{y} \alpha p_{Y}^1(y)+\max_{y}(1-\alpha)p_Y^2(y)\\
&=\alpha\max_{y} p_{Y}^1(y)+(1-\alpha)\max_{y}p_Y^2(y),
\end{align*}
which results in $-\max_{y}p_{Y}(y)$ being a concave functional of $p_Y(\cdot)$. Hence, following Proposition 1, the problem reduces to 
\begin{align}
E_{\epsilon}(X,Y)=1+\!\!\!\!\!\!\!\!\!\!\!\!\!\!\!\!\!\!\!\!\!\!\!\!\!\!\!\!\!&\min_{\substack{\mathbf{w}\geq 0:\\\begin{bmatrix}
f(\mathbf{s}_1)&f(\mathbf{s}_2)&\dots&f(\mathbf{s}_K)
\end{bmatrix}\cdot\mathbf{w}\leq2\epsilon \\\begin{bmatrix}
\mathbf{s}_1&\mathbf{s}_2&\dots&\mathbf{s}_K
\end{bmatrix}\cdot\mathbf{w}=\mathbf{p}_Y}}\ \!\!\!\!\!\!\!\!\!\!\!\!\!\!\!\!\!\!\!\!\!\!\!\!\!\!\!\!\!-\begin{bmatrix}
s_{m_1}&s_{m_2}&\dots&s_{m_K}
\end{bmatrix}\cdot\mathbf{w}
,\label{ewq}
\end{align}
where $s_{m_i}$ is the maximum element of the vector $\mathbf{s}_i,\ i\in[1:K].$ Once the LP is solved, 
if $w_i^*\neq 0$ ($i\in[1:K]$), the value of $u_i$ is set as the maximum element of the probability vector $\mathbf{p}_{Y|u_i}=\mathbf{s}_i$. 

Similarly to Theorem 1, it can be verified that when $Y$ is binary, the problem in (\ref{errorp}) has a closed form solution given by
{\color{black}
\begin{align}
E_{\epsilon}(X,Y)&=\min\{p,1-p\}\bigg(\!1-\frac{\epsilon}{p(1-p)\|\mathbf{p}_{X|y_1}-\mathbf{p}_{X|y_2}\|_1}\!\bigg)^+\!.
\end{align}
}
\section{}\label{app1}
Let $\mathbf{p}_{Y|u}$ be given as $\mathbf{p}_{Y|u}=\lambda \mathbf{p}_{Y|u_1}+(1-\lambda)\mathbf{p}_{Y|u_2}$, where $\lambda\in[0,1]$. It is obvious that for an arbitrary function $b(\cdot)$,
\begin{equation}\label{fu}
\mathds{E}[b(Y)|U=u]=\lambda\mathds{E}[b(Y)|U=u_1]+(1-\lambda)\mathds{E}[b(Y)|U=u_2].
\end{equation}
Therefore,
\begin{align}
\mbox{Var}[Y|U=u]&=\mathds{E}\bigg[\bigg(Y-\mathds{E}[Y|U=u]\bigg)^2\bigg|U=u\bigg]\nonumber\\
&=\mathds{E}[Y^2|U=u]-\bigg(\mathds{E}[Y|U=u]\bigg)^2\nonumber\\
&=\lambda\mathds{E}[Y^2|U=u_1]+(1-\lambda)\mathds{E}[Y^2|U=u_2]\nonumber\\
&\ \ \ -\bigg(\lambda\mathds{E}[Y|U=u_1]+(1-\lambda)\mathds{E}[Y|U=u_2]\bigg)^2\label{c0}\\
&\geq \lambda\mathds{E}[Y^2|U=u_1]+(1-\lambda)\mathds{E}[Y^2|U=u_2]\nonumber\\
&\ \ \ -\lambda\bigg(\mathds{E}[Y|U=u_1]\bigg)^2-(1-\lambda)\bigg(\mathds{E}[Y|U=u_2]\bigg)^2\label{co}\\
&=\lambda\mathds{E}\bigg[\bigg(Y-\mathds{E}[Y|U=u_1]\bigg)^2\bigg|U=u_1\bigg]\nonumber\\
&\ \ \ +(1-\lambda)\mathds{E}\bigg[\bigg(Y-\mathds{E}[Y|U=u_2]\bigg)^2\bigg|U=u_2\bigg]\nonumber
\end{align}
\begin{align}
&=\lambda\mbox{Var}[Y|U=u_1]+(1-\lambda)\mbox{Var}[Y|U=u_2]\nonumber,
\end{align}
where (\ref{c0}) follows from (\ref{fu}); and (\ref{co}) is due to the convexity of $x^2$. 
{\color{black}
\section{Proof of theorem 5}\label{app1.5}
We have
\begin{align}
I(X;U)&=\mathds{E}_U[D(p_{X|U}(\cdot|U)||p_X(\cdot))]\nonumber\\
&\geq \mathds{E}_U[2\log_2e\cdot \delta^2(p_{X|U}(\cdot|U),p_X(\cdot))]\label{pins}\\
&\geq 2\log_2e\bigg(\mathds{E}_U[\delta(p_{X|U}(\cdot|U),p_X(\cdot))]\bigg)^2\label{x22}\\
&=2\log_2e\cdot T^2(X;U)\nonumber,
\end{align}
where (\ref{pins}) comes from the application of Pinsker's inequality, and (\ref{x22}) follows from the convexity of $x^2$ in $x$ and Jensen's inequality.

For (\ref{bound2}), we proceed as follows. From (\ref{maxl}) and its following explanation on $\mathcal{X}$, maximal leakage can be rewritten as
\begin{align}
\mathcal{L}(X\to U)&=\log\sum_{u\in\mathcal{U}}p_U(u)\max_{x}\frac{p_{X|U}(x|u)}{p_X(x)}\label{maxl2}
\end{align}
For an arbitrary pmf $q_X(\cdot)$ on $\mathcal{X}$, it can be verified that
\footnote{This can be proved by contradiction. Assume that $\exists x_0\in\mathcal{X}$ such that (\ref{seyek}) does not hold. As a result
\begin{align*}
q_X(x_0)- p_{X}(x_0)&>\sum_{x\neq x_0}|q_X(x)- p_{X}(x)|\\
&\geq\sum_{x\neq x_0}p_{X}(x)-q_X(x)\\
&=q_X(x_0)- p_{X}(x_0),
\end{align*}
which is a contradiction.
}
\begin{equation}\label{seyek}
q_X(x)\leq p_{X}(x)+\frac{1}{2}\|\mathbf{q}_{X}-\mathbf{p}_X\|_1,\ \forall x\in\mathcal{X}.
\end{equation}
Therefore,
\begin{align}
\max_{x}\frac{q_{X}(x)}{p_X(x)}&\leq\max_{x}\frac{p_{X}(x)+\frac{1}{2}\|\mathbf{q}_{X}-\mathbf{p}_X\|_1}{p_X(x)}\label{3sd}\\&
=\frac{\min_xp_X(x)+\frac{1}{2}\|\mathbf{q}_{X}-\mathbf{p}_X\|_1}{\min_xp_X(x)}\label{akhj},
\end{align}
where (\ref{3sd}) follows from (\ref{seyek}), and (\ref{akhj}) from the fact that for $a,t>0$, $\frac{t+a}{t}$ is strictly decreasing in $t$. Replacing $q_X(\cdot)$ with $p_{X|U}(\cdot|u)$ in (\ref{3sd}) and (\ref{akhj}), and plugging the result into (\ref{maxl2}) results in (\ref{bound2}).

The inequality in (\ref{bound3}) is proved as follows. Let $\Delta_x\triangleq q_X(x)-p_X(x),\ \forall x\in\mathcal{X}$. Hence, we have $\sum_{x\in\mathcal{X}}\Delta_x=0$. Define
\begin{equation*}
\mathcal{X}^+\triangleq\{x\in\mathcal{X}|\Delta_x\geq 0\},\ \mathcal{X}^-\triangleq\mathcal{X}\backslash\mathcal{X}^+.
\end{equation*}
Therefore, we can write
\begin{align}
\max_{x}\frac{q_{X}(x)}{p_X(x)}&=\max_{x}\frac{p_{X}(x)+\Delta_x}{p_X(x)}\nonumber\\
&=1+\max_{x\in\mathcal{X}^+}\frac{\Delta_x}{p_X(x)}\label{asq1}\\
&\geq 1+\frac{\max_{x\in\mathcal{X}^+}\Delta_x}{\max_{x\in\mathcal{X}}p_X(x)}\nonumber\\
&\geq 1+\frac{\frac{1}{2}\|\mathbf{q}_{X}-\mathbf{p}_X\|_1}{|\mathcal{X}^+|\max_{x\in\mathcal{X}}p_X(x)}\label{asq2}\\
&\geq 1+\frac{\frac{1}{2}\|\mathbf{q}_{X}-\mathbf{p}_X\|_1}{(|\mathcal{X}|-1)\max_{x\in\mathcal{X}}p_X(x)}\label{asq3},
\end{align}
where (\ref{asq1}) follows from the definition of $\mathcal{X}^+$; (\ref{asq2}) follows from the fact that
\begin{equation*}
\sum_{x\in\mathcal{X}^+}\Delta_x=\frac{1}{2}\|\mathbf{q}_{X}-\mathbf{p}_X\|_1,
\end{equation*}
and the maximum values for $\Delta_x$ is minimized when all of them are equal. When $\mathbf{q}_{X}=\mathbf{p}_X$, (\ref{asq3}) is obvious, and when $\mathbf{q}_{X}\neq\mathbf{p}_X$, we have $|\mathcal{X}^+|<|\mathcal{X}|$, and (\ref{asq3}) holds. Finally, Replacing $q_X(\cdot)$ with $p_{X|U}(\cdot|u)$, and using (\ref{maxl2}) results in (\ref{bound3}).
}
\section{}\label{app2}
The procedure of finding the elements of $\mathbb{S}$ is as follows.
We can write $\mathbb{S}_i=\{\mathbf{x}\in\mathbb{R}^{|\mathcal{Y}|}|\tilde{\mathbf{A}}_i\mathbf{x}\leq{\mathbf{b}_i},\mathbf{1}_{|\mathcal{Y}|}^T\cdot\mathbf{x}=1,\mathbf{x}\geq 0\}$. Matrix $\tilde{\mathbf{A}}_i$ has $|\mathcal{Y}|$ columns and at least two (at most $|\mathcal{X}|$) rows that correspond to the sign determination of the elements in the $L^1$-norm. The extreme points of $\mathbb{S}_i$ are obtained from the basic feasible solutions (see \cite{LP1}, \cite{LP2}) of their corresponding set denoted by $\mathbb{D}_i=\{\mathbf{x}\in\mathbb{R}^{|\mathcal{Y}|'}|\mathbf{A}_i\mathbf{x}=\mathbf{b}_i,\mathbf{x}\geq 0\}$. These corresponding sets are obtained by adding slack variables to change the inequality constraints of $\mathbb{S}_i$ into equality. Matrix $\mathbf{A}_i$ has at most $|\mathcal{X}|+1$ rows (taking into account $\mathbf{1}_{|\mathcal{Y}|}^T\cdot\mathbf{x}=1$). 

The procedure of finding the basic feasible solutions of $\mathbb{D}_i$ is as follows. Let $r_i$ denote the number of rows in $\mathbf{A}_i$. Pick a set $\mathcal{B}\subset[1:|\mathcal{Y}|']$ of indices that correspond to $r_i$ linearly independent columns of matrix $\mathbf{A}_i$. There are at most $|\mathcal{Y}|'\choose {r_i}$ ways of choosing $r_i$ linearly independent columns of $\mathbf{A}_i$. Let $\mathbf{A}_{\mathcal{B}}$ be an $r_i\times r_i$ matrix whose columns are the columns of $\mathbf{A}_i$ indexed by the indices in $\mathcal{B}$. Also, for any $\mathbf x\in\mathbb{D}$, let $\tilde{\mathbf x}=\begin{bmatrix}\mathbf{x}_{\mathcal{B}}^T&\mathbf{x}_{\mathcal{N}}^T\end{bmatrix}^T$, where $\mathbf{x}_{\mathcal{B}}$ and $\mathbf{x}_{\mathcal{N}}$ are $r_i$-dimensional and $(|\mathcal{Y}|'-r_i)$-dimensional vectors whose elements are the elements of $\mathbf{x}$ indexed by the indices in $\mathcal{B}$ and $[1:|\mathcal{Y}|']\backslash\mathcal{B}$, respectively.


For any basic feasible solution $\mathbf{x}^*$, there exists a set $\mathcal{B}\subset[1:|\mathcal{Y}|']$ of indices that correspond to a set of linearly independent columns of $\mathbf{A}_i$, such that the corresponding vector of $\mathbf{x}^*$, i.e., $\tilde{\mathbf{x}}^*=\begin{bmatrix}{\mathbf{x}^*_\mathcal{B}}^T&{\mathbf{x}^*_\mathcal{N}}^T\end{bmatrix}^T$, satisfies the following
\begin{equation*}
\mathbf{x}_\mathcal{N}^*=\mathbf{0},\ \ \ \mathbf{x}_\mathcal{B}^*=\mathbf{A}_\mathcal{B}^{-1}\mathbf{b},\ \ \ \mathbf{x}_\mathcal{B}^*\geq 0.
\end{equation*}
On the other hand, for any set $\mathcal{B}\subset[1:|\mathcal{Y}|']$ of indices that correspond to a set of linearly independent columns of $\mathbf{A}_i$, if $\mathbf{A}_\mathcal{B}^{-1}\mathbf{b}\geq 0$, then $\begin{bmatrix}
\mathbf{A}_\mathcal{B}^{-1}\mathbf{b}\\\mathbf{0}
\end{bmatrix}$ is the corresponding vector of a basic feasible solution.
Hence, the basic feasible solutions of $\mathbb{D}_i$ can be obtained in this way. 

As an example consider the joint distribution shown in Figure \ref{fig3}, where $\mathbf{p}_X=\mathbf{p}_Y=\begin{bmatrix}
\frac{1}{3}&\frac{1}{3}&\frac{1}{3}
\end{bmatrix}^T$ and the elements of the transition matrix $\mathbf{P}_{X|Y}$ are shown in the figure.
From (\ref{efraz}), we have
\begin{align*}
f(\mathbf{x})&=\frac{1}{2}\|\mathbf{P}_{X|Y}\left(\mathbf{x}-\mathbf{p}_Y\right)\|_1\\
&=\frac{1}{3}\bigg(|2x_1+x_2-1|+|x_2+2x_3-1|\bigg)
\end{align*}
%
The sign determination of the absolute value terms results in the four possible regions given by 
\begin{equation}
\mathbb{S}_i=\{\mathbf{x}\in\mathbb{R}^{3}|\tilde{\mathbf{A}}_i\mathbf{x}\leq{\mathbf{b}_i},\mathbf{1}_{3}^T\cdot\mathbf{x}=1,\mathbf{x}\geq 0\},\forall i\in[1:4],
\end{equation}
where
\begin{equation*}
\tilde{\mathbf{A}}_1=\begin{bmatrix}
-2&-1&0\\0&-1&-2
\end{bmatrix},\mathbf{b}_1=\begin{bmatrix}
-1\\-1
\end{bmatrix},\tilde{\mathbf{A}}_2=\begin{bmatrix}
2&1&0\\0&1&2
\end{bmatrix},
\end{equation*}
\begin{equation*}
\mathbf{b}_2=\begin{bmatrix}
1\\1
\end{bmatrix},\tilde{\mathbf{A}}_3=\begin{bmatrix}
-2&-1&0\\0&1&2
\end{bmatrix},\mathbf{b}_3=\begin{bmatrix}
-1\\1
\end{bmatrix},
\end{equation*}
\begin{equation*}
\tilde{\mathbf{A}}_4=\begin{bmatrix}
2&1&0\\0&-1&-2
\end{bmatrix},\mathbf{b}_4=\begin{bmatrix}
1\\-1
\end{bmatrix}.
\end{equation*}
In order to find the extreme points of ${\mathbb{S}}_1$, we need to introduce the slack variables $x_4,x_5\geq 0$ to change the two inequality constraints of ${\mathbb{S}}_1$ into equality. As a result, we have the following set
\begin{equation*}
\mathbb{D}_1=\bigg\{\mathbf{x}\in\mathbb{R}^5\bigg|\mathbf{A}_1\mathbf{x}=\mathbf{b}_1,\mathbf{x}\geq 0\bigg\},
\end{equation*}
where
\begin{equation*}
\mathbf{A}_1=\begin{bmatrix}
-2&-1&0&1&0\\0&-1&-2&0&1\\1&1&1&0&0
\end{bmatrix},\mathbf{b}_1=\begin{bmatrix}
1\\1\\1
\end{bmatrix}
\end{equation*}
In order to obtain the basic feasible solutions of $\mathbb{D}_1$, we observe that there are at most $5\choose 3$ ways of choosing 3 linearly independent columns of $\mathbf{A}_1$. Excluding the index set $\{1,2,3\}$, as the columns corresponding to this index set are linearly dependent, $\mathcal{B}$ can be any of $\{1,2,4\},\{1,2,5\},\{1,3,4\},\{1,3,5\},\{1,4,5\},$ $\{2,3,4\},\{2,3,5\},\{2,4,5\}$, and $\{3,4,5\}$. By obtaining the values of $\mathbf{x}_{\mathcal{B}}=\mathbf{A}_{1_{\mathcal{B}}}^{-1}\mathbf{b}_1$ corresponding to these 9 possibilities, and checking their feasibility condition $\mathbf{x}_{\mathcal{B}}\geq 0$, we conclude\footnote{A much easier way to obtain the extreme points of ${\mathbb{S}}_1$ (and also ${\mathbb{S}}_2$) in this example is by noting that ${\mathbb{S}}_1$ (${\mathbb{S}}_2$) is a straight line between the two points $\begin{bmatrix}0&1&0\end{bmatrix}^T$ and $\begin{bmatrix}\frac{1}{2}&0&\frac{1}{2}\end{bmatrix}^T$. Nonetheless, we treated it as a general region to show the procedure of finding the extreme points.} that the extreme points of ${\mathbb{S}}_1$ are $\begin{bmatrix}0&1&0\end{bmatrix}^T$ and $\begin{bmatrix}\frac{1}{2}&0&\frac{1}{2}\end{bmatrix}^T$. In a similar way, the extreme points of the regions ${\mathbb{S}}_2$ to ${\mathbb{S}}_4$ can be obtained. 
\bibliography{REFERENCE}

\begin{thebibliography}{10}
\providecommand{\url}[1]{#1}
\csname url@samestyle\endcsname
\providecommand{\newblock}{\relax}
\providecommand{\bibinfo}[2]{#2}
\providecommand{\BIBentrySTDinterwordspacing}{\spaceskip=0pt\relax}
\providecommand{\BIBentryALTinterwordstretchfactor}{4}
\providecommand{\BIBentryALTinterwordspacing}{\spaceskip=\fontdimen2\font plus
\BIBentryALTinterwordstretchfactor\fontdimen3\font minus
  \fontdimen4\font\relax}
\providecommand{\BIBforeignlanguage}[2]{{%
\expandafter\ifx\csname l@#1\endcsname\relax
\typeout{** WARNING: IEEEtran.bst: No hyphenation pattern has been}%
\typeout{** loaded for the language `#1'. Using the pattern for}%
\typeout{** the default language instead.}%
\else
\language=\csname l@#1\endcsname
\fi
#2}}
\providecommand{\BIBdecl}{\relax}
\BIBdecl

\bibitem{Rebollo}
D.~Rebollo-Monedero, J.~Forne, and J.~Domingo-Ferrer, ``From t-closeness-like
  privacy to postrandomization via information theory,'' \emph{IEEE
  Transactions on Knowledge and Data Engineering}, vol.~22, no.~11, pp.
  1623--1636, Nov 2010.

\bibitem{Rebollo2}
D.~Rebollo-Monedero and J.~Forne, ``Optimized query forgery for private
  information retrieval,'' \emph{IEEE Transactions on Information Theory},
  vol.~56, no.~9, pp. 4631--4642, Sept 2010.

\bibitem{Sankar3}
L.~Sankar, S.~R. Rajagopalan, and H.~V. Poor, ``Utility-privacy tradeoffs in
  databases: An information-theoretic approach,'' \emph{IEEE Transactions on
  Information Forensics and Security}, vol.~8, no.~6, pp. 838--852, June 2013.

\bibitem{Sankar5}
L.~Sankar, S.~R. Rajagopalan, S.~Mohajer, and S.~Mohajer, ``Smart meter
  privacy: A theoretical framework,'' \emph{IEEE Transactions on Smart Grid},
  vol.~4, no.~2, pp. 837--846, June 2013.

\bibitem{Khisti2}
S.~Li, A.~Khisti, and A.~Mahajan, ``Information-theoretic privacy for smart
  metering systems with a rechargeable battery,'' \emph{IEEE Transactions on
  Information Theory}, vol.~64, no.~5, pp. 3679--3695, May 2018.

\bibitem{Tan}
O.~Tan, D.~Gunduz, and H.~V. Poor, ``Increasing smart meter privacy through
  energy harvesting and storage devices,'' \emph{IEEE Journal on Selected Areas
  in Communications}, vol.~31, no.~7, pp. 1331--1341, 2013.

\bibitem{Giulio}
G.~Giaconi, D.~Gündüz, and H.~V. Poor, ``Smart meter privacy with renewable
  energy and an energy storage device,'' \emph{IEEE Transactions on Information
  Forensics and Security}, vol.~13, no.~1, pp. 129--142, 2018.

\bibitem{Calmon1}
F.~Calmon and N.~Fawaz, ``Privacy against statistical inference,'' in
  \emph{50th Annual Allerton Conference}, Illinois, USA, Oct. 2012, pp.
  1401--1407.

\bibitem{Makhdoumi}
A.~Makhdoumi, S.~Salamatian, N.~Fawaz, and M.~M\'{e}dard, ``From the
  information bottleneck to the privacy funnel,'' in \emph{IEEE Information
  Theory Workshop (ITW)}, 2014, pp. 501--505.

\bibitem{Shahab5}
S.~Asoodeh, M.~Diaz, F.~Alajaji, and T.~Linder, ``Estimation efficiency under
  privacy constraints,'' \emph{IEEE Transactions on Information Theory}, pp.
  1--1, 2018.

\bibitem{As1}
------, ``Privacy-aware guessing efficiency,'' in \emph{2017 IEEE International
  Symposium on Information Theory (ISIT)}, June 2017, pp. 754--758.

\bibitem{Ishwar}
Y.~Wang, Y.~Basciftci, and P.~Ishwar, ``Privacy-utility tradeoffs under
  constrained data release mechanisms,''
  \emph{https://arxiv.org/pdf/1710.09295.pdf}.

\bibitem{Shahab6}
S.~Asoodeh, F.~Alajaji, and T.~Linder, ``Privacy-aware {MMSE} estimation,'' in
  \emph{IEEE International Symposium on Information Theorty (ISIT)}, 2016, pp.
  1989--1993.

\bibitem{Huang1}
C.~Huang, P.~Kairouz, X.~Chen, L.~Sankar, and R.~Rajagopal, ``Context-aware
  generative adversarial privacy,'' \emph{Entropy}, vol.~19, no.~12, 2017.

\bibitem{SA2}
S.~Asoodeh, M.~Diaz, F.~Alajaji, and T.~Linder, ``Information extraction under
  privacy constraints,'' \emph{Information}, vol.~7, no.~1, 2016.

\bibitem{Ish22}
A.~Tripathy, Y.~Wang, and P.~Ishwar, ``Privacy-preserving adversarial
  networks,'' \emph{CoRR}, vol. abs/1712.07008, 2017.

\bibitem{Tishby}
N.~Tishby, F.~Pereira, and W.~Bialek, ``The information bottleneck method,'' in
  \emph{37th Annual Allerton Conference on Communication, Control and
  Computing}, 2000, pp. 368--377.

\bibitem{Bori}
B.~Rassouli and D.~Gunduz, ``Optimal utility-privacy trade-off with total
  variation distance as a privacy measure,'' in \emph{to appear in the IEEE
  Information Theory Workshop (ITW)}, 2018.

\bibitem{Issa}
I.~Issa, S.~Kamath, and A.~Wagner, ``An operational measure of information
  leakage,'' in \emph{Inf. Sci. and Sys. (CISS)}, 2016, pp. 234--239.

\bibitem{Liao}
J.~Liao, O.~Kosut, L.~Sankar, and F.~P. Calmon, ``A tunable measure for
  information leakage,'' in \emph{2018 IEEE International Symposium on
  Information Theory (ISIT)}, June 2018, pp. 701--705.

\bibitem{Shahab1}
S.~Asoodeh, F.~Alajaji, and T.~Linder, ``Notes on information-theoretic
  privacy,'' in \emph{52nd Annual Allerton Conference}, Illinois, USA, Oct.
  2014, pp. 1272--1278.

\bibitem{Wang2}
H.~Wang and F.~P. Calmon, ``An estimation-theoretic view of privacy,'' in
  \emph{2017 55th Annual Allerton Conference on Communication, Control, and
  Computing (Allerton)}, Oct 2017, pp. 886--893.

\bibitem{chenchen}
J.~Chen, J.~Konrad, and P.~Ishwar, ``Vgan-based image representation learning
  for privacy-preserving facial expression recognition,'' \emph{CoRR}, vol.
  abs/1803.07100, 2018.

\bibitem{NIPS}
I.~Goodfellow, J.~Pouget-Abadie, M.~Mirza, B.~Xu, D.~Warde-Farley, S.~Ozair,
  A.~Courville, and Y.~Bengio, ``Generative adversarial nets,'' in
  \emph{Advances in Neural Information Processing Systems 27}.\hskip 1em plus
  0.5em minus 0.4em\relax Curran Associates, Inc., 2014, pp. 2672--2680.

\bibitem{Wang3}
H.~Wang, M.~Diaz, F.~P. Calmon, and L.~Sankar, ``The utility cost of robust
  privacy guarantees,'' in \emph{2018 IEEE International Symposium on
  Information Theory (ISIT)}, June 2018, pp. 706--710.

\bibitem{Elgamal}
A.~E. Gamal and Y.-H. Kim, \emph{Network Information Theory}.\hskip 1em plus
  0.5em minus 0.4em\relax Cambridge University Press, 2012.

\bibitem{Levy}
B.~C. Levy, \emph{Principles of Signal Detection and Parameter
  Estimation}.\hskip 1em plus 0.5em minus 0.4em\relax Springer, 2008.

\bibitem{LP1}
D.~Bertsimas and J.~N. Tsitsiklis, \emph{Introduction to linear
  optimization}.\hskip 1em plus 0.5em minus 0.4em\relax Athena Scientic, 1997.

\bibitem{LP2}
K.~G. Murty, \emph{Linear Programming}.\hskip 1em plus 0.5em minus 0.4em\relax
  John Wiley and Sons, 1983.

\end{thebibliography}
\bibliographystyle{IEEEtran}
\vspace{-1cm}
\begin{IEEEbiography}[{\includegraphics[width=1in,height=1.25in,clip,keepaspectratio]{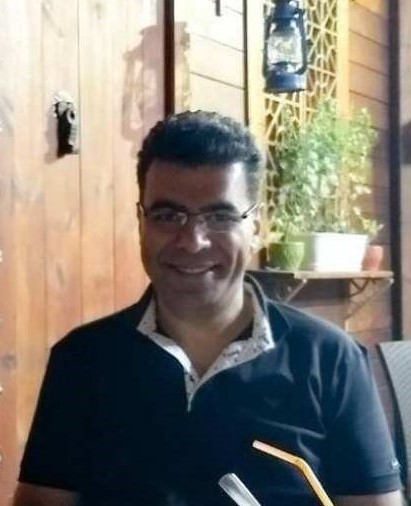}}]{Borzoo Rassouli}
received the M.Sc. degree in electrical engineering from university of Tehran, Iran in 2012, and the Ph.D. degree in communications engineering from Imperial College London, UK in 2016. He was a postdoctoral research associate at Imperial College from 2016 to 2018. In August 2018, he joined university of Essex as a lecturer (Assistant Professor). His research interests lie in the general areas of information theory and statistics.
\end{IEEEbiography}
\vfill
\vspace{-7cm}
\begin{IEEEbiography}[{\includegraphics[width=1in,height=1.25in,clip,keepaspectratio]{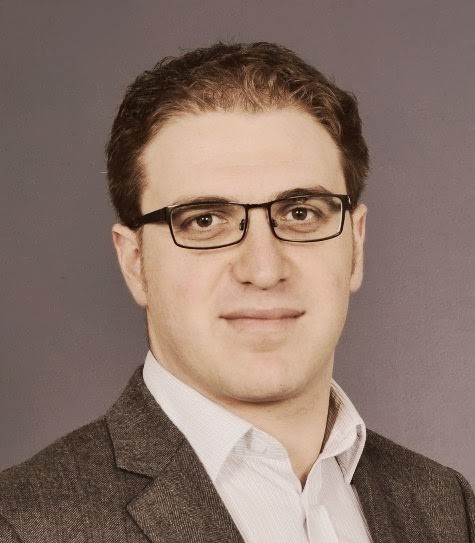}}]{Deniz G\"{u}nd\"{u}z}
[S’03-M’08-SM’13] received the M.S. and Ph.D. degrees in electrical engineering from NYU Tandon School of Engineering (formerly Polytechnic University) in 2004 and 2007, respectively. After his PhD, he served as a postdoctoral research associate at Princeton University, and as a consulting assistant professor at Stanford University. He was a research associate at CTTC in Barcelona, Spain until September 2012, when he joined the Electrical and Electronic Engineering Department of Imperial College London, UK, where he is currently a Reader (Associate Professor) in information theory and communications, and leads the Information Processing and Communications Lab.
His research interests lie in the areas of communications and information theory, machine learning, and privacy. Dr. Gündüz is an Editor of the IEEE Transactions on Green Communications and Networking, and a Guest Editor of the IEEE Journal on Selected Areas in Communications, Special Issue on Machine Learning in Wireless Communication. He is the recipient of the IEEE Communications Society - Communication Theory Technical Committee (CTTC) Early Achievement Award in 2017, a Starting Grant of the European Research Council (ERC) in 2016, IEEE Communications Society Best Young Researcher Award for the EMEA Region in 2014, Best Paper Award at the 2016 IEEE WCNC, and the Best Student Paper Awards at the 2018 IEEE WCNC and the 2007 IEEE ISIT.
\end{IEEEbiography}
\vfill
\end{document}